\documentclass[twocolumn,preprintnumbers,amsmath,amssymb,amsfonts]{revtex4}

\usepackage{graphicx}% Include figure files
\usepackage{mathrsfs}
\usepackage{amssymb}
\usepackage{color}
\usepackage{ulem}
\usepackage{mathtools}
\usepackage{float}
\usepackage{multirow}
\usepackage{diagbox}

\begin{document}
	\title{Fisher discord as a quantifier of quantum complexity}
	\author{Huihui Li}
    \affiliation{Yau Mathematical Sciences Center, Tsinghua University, Beijing 100084, China}
	\author{Shunlong Luo and Yue Zhang}
	 \email{zhangyue115@amss.ac.cn}
	\affiliation{State Key Laboratory of Mathematical Sciences, Academy of Mathematics and
		Systems Science, Chinese Academy of Sciences,  Beijing 100190, China\\
		School of Mathematical Sciences, University of Chinese
		Academy of Sciences, Beijing 100049, China}
	
\begin{abstract}
Two classically equivalent expressions of mutual information of probability distributions (classical bipartite states) diverge when extended  to quantum systems, and this difference has been employed to define quantum discord, a quantifier of quantum correlations beyond entanglement. Similarly, equivalent expressions of classical Fisher information of parameterized probability distributions diverge when extended to quantum states, and this difference may be exploited to characterize the complex nature of quantum states. By complexity of quantum states, we mean some hybrid nature which intermingles the classical and quantum features. It is desirable to quantify complexity of quantum states from various perspectives. In this work, we pursue the idea of discord and  introduce an information-theoretic quantifier of complexity for quantum states (relative to the  Hamiltonian that drives the evolution of quantum systems) via the notion of Fisher discord, which is  defined by the difference between two important versions of quantum Fisher information: the quantum Fisher information defined via the symmetric logarithmic derivatives and  the Wigner-Yanase skew information defined via the square roots of quantum states. We reveal basic properties of the quantifier of complexity, and compare it with some other quantifiers of
complexity. In particular, we show that equilibrium states (or stable states, which commute with the Hamiltonian of the quantum system) and all pure states exhibit zero complexity in this setting. As illustrations, we evaluate the complexity for various prototypical states in both discrete and continuous-variable quantum systems.

\vskip 0.3cm
\noindent {\bf Keywords}: Classical Fisher information, quantum Fisher information, Wigner-Yanase skew information, Fisher discord, quantum complexity
		
\vskip 0.3cm
		
		%\noindent {\bf PACS numbers}: 03.65.Ta, 03.65Wj, 42.50Dv
%\pacs{03.65.Ta, 03.67.-a, 42.50-p}
		
	\end{abstract}
           %The generalization of the  quantifier of complexity based on  metric adjusted skew information has been discussed.

          %It may shed light on quantum metrology.

	\maketitle
	
\section{Introduction}

 %stable
Complexity is a multifaceted  and ubiquitous concept that quantifies the intricacy  of systems  and their divergence from idealization or simplification.  It holds prominence in numerous  realms of science \cite{suh2005complexity,battiston2016complexity,luo2012regularity,papadimitriou2003computational,nicolis1989exploring,parisi1993statistical,gell2002complexity,ruiz2013statistical,wigderson2006p}, such as engineering \cite{suh2005complexity,luo2012regularity}, financial \cite{battiston2016complexity}, computer science \cite{papadimitriou2003computational}, physics \cite{nicolis1989exploring,parisi1993statistical,gell2002complexity,ruiz2013statistical},  and  mathematics \cite{wigderson2006p}, and so on.

In  quantum information theory,  complexity is crucial for understanding the interplay between classicality and quantumness \cite{bernstein1993quantum,watrous2012quantum}.
A variety of quantifiers of complexity have been extensively investigated and applied \cite{mccabe1976complexity,yi2024complexity,mora2006algorithmic,jefferson2017circuit,LOPEZRUIZ1995321,calbet2001tendency,catalan2002features,romera2004fisher,tang2025quantifying,ohya1998foundation,shiner1999simple,valdez2017quantifying,lloyd2001measures}, such as  computational complexity \cite{mccabe1976complexity,yi2024complexity}, algorithmic complexity \cite{mora2006algorithmic}, circuit complexity \cite{jefferson2017circuit}, statistical complexity  \cite{LOPEZRUIZ1995321,ohya1998foundation,shiner1999simple,calbet2001tendency,catalan2002features,romera2004fisher}, phase-space complexity \cite{tang2025quantifying}, quantum phase transitions complexity \cite{valdez2017quantifying},  and so on \cite{lloyd2001measures}.
% entropy  \cite{ohya1998foundation,shiner1999simple}, and mutual information complex networks \cite{valdez2017quantifying}
Each quantifier of complexity offers unique perspectives. Complexity may carry different meanings in different contexts. In this work, we will regard a quantum state as simple if it is a quantum pure  state, or a purely classical state in some sense, and regard it as possessing some complexity if it mixes both classical and quantum features. Given the intricate nature of complexity, it is desirable to characterize and quantify complexity in the quantum context from a wide range of angles.

Originated from statistical inference, the fundamental notion of Fisher information was first proposed by Fisher in 1925 as a quantifier of the information content of an observable random variable about an unknown parameter in a statistical model \cite{Fish1925,Cram1946}. Its quantum analogue, the quantum Fisher information,  plays a crucial role in quantum metrology due to the  celebrated Cram\'er-Rao inequality in quantum parameter estimation
\cite{Wign1963,Hels1969,Hole1973,Petz1996,Brau1994,hansen2008metric,luo2003wigner,luo2004wigner,sun2022quantifying,Liu2019,Lu2013,holevo2011probabilistic,liu2020quantum,Scan2024}. Recent studies have revealed that quantum Fisher information is closely linked to other aspects of quantum mechanics, such as  entanglement \cite{hyllus2012fisher,li2013entanglement}, quantum uncertainty relation \cite{luo2000quantum,gibilisco2007uncertainty,toth2022uncertainty},  and quantum non-Markovianity \cite{song2015quantum}, and quantum coherence \cite{feng2017quantifying,sun2023coherence}.

Unlike classical Fisher information,  quantum Fisher information  is not unique  due to the intrinsic noncommutative nature of quantum mechanics. Various versions of quantum Fisher information have been introduced ever since 1960s \cite{Wign1963,Hels1969,Hole1973,Petz1996,Brau1994,hansen2008metric,luo2003wigner,luo2004wigner}.  The Wigner-Yanase skew information is a distinguished version of quantum Fisher information  with respect to the time parameter encoded in the evolution of quantum state driven by the conserved quantity (Hamiltonian) \cite{luo2003wigner,luo2004wigner}, which was originally introduced by Wigner and Yanase \cite{Wign1963}.
% to quantify the information content of a quantum state relative to observables non-commuting with the conserved observable in the context of quantum measurement
The skew information exhibits numerous notable properties and has become increasingly significant in  quantum information theory \cite{Lieb1973,wehrl1978general,connes1978homogeneity}. It is a pivotal tool for characterizing quantum resources \cite{luo2017quantum,luo2018coherence,luo2012quantifying,Sun2021,luo2005heisenberg,luo2006quantum,luo2019quantifying,li2023probing,Li2025,Zhan2021}, such as coherence \cite{luo2017quantum,luo2018coherence,Zhan2021},  correlations \cite{luo2012quantifying,li2023probing,Li2025}, asymmetry \cite{Sun2021}, uncertainty relation \cite{luo2005heisenberg,luo2006quantum}, and nonclassicality \cite{luo2019quantifying}.

In this work, motivated by the idea of quantum discord \cite{ollivier2001quantum,luo2008quantum}, which is defined by the difference between two quantum natural extensions of equivalent expressions of the classical mutual information, and inspired by the non-uniqueness of quantum Fisher information as quantum extensions of the classical Fisher information \cite{li2016fisher,shun2006fisher}, we introduce an information-theoretic quantifier of complexity for quantum states via Fisher discord, defined as the difference between prominent versions of quantum Fisher information: One is defined via symmetric logarithmic derivative, and the other is the Wigner-Yanase skew information defined via square root of quantum states.

The remainder of the work is organized as follows: In Sec. II, we review two versions of quantum Fisher information, i.e., the quantum Fisher information based on the symmetric logarithmic derivative and the Wigner-Yanase skew information. We then propose the difference between them as a quantifier of  complexity and investigate its properties in Sec. III. In Sec. IV, we evaluate it for paradigmatic Hamiltonians, and illustrate its utility in quantifying the complexity of qubit states and single-mode Bosonic  states.   We perform  a comprehensive comparative analysis of our complexity quantifier against several existing quantifiers of complexity in Sec. V.    Finally,  we  conclude with a summary in Sec. VI.  
%We calculate the complexity relative to  a versatile  Hamiltonian for quantum optical systems in the Appendix.

\section{Quantum Fisher information}

Let us review the mathematical definition of Fisher information and its two natural extensions to the quantum context. As outlined in Refs. \cite{Fish1925,Cram1946}, the Fisher information for a parametric  classical probability distribution $p_{\theta}$ is defined as \cite{Fish1925,Cram1946,luo2004wigner}
\begin{align}
I_{\rm F}(p_{\theta}) &= \int\Big(\frac{\rm \partial}{{\rm \partial} \theta}\ln{p_{\theta}(x)}\Big)^2p_{\theta}(x){\rm d}x,\label{D1}\\
&=4\int\Big(\frac{\rm \partial}{{\rm \partial} \theta}\sqrt{p_{\theta}(x)}\Big)^2{\rm d}x.\label{D2}
\end{align}

To extend classical Fisher information to quantum systems, parametric probability distributions $ p_\theta $ are replaced with parametric density operators $ \rho_\theta $, and integration is substituted with the trace operation. Inspired by the following trivial identity
\begin{align}
\frac{\rm \partial}{{\rm \partial} \theta}{p_\theta}=\frac{1}{2}(l_{\theta}p_{\theta}+p_{\theta}l_{\theta}) \label{logd}
\end{align}
for the statistical  score function $l_{\theta}={\rm \partial}\ln{p_\theta} /{{\rm \partial} \theta}$ in Eq. (\ref{D1}), Helstrom \cite{Hels1969} first proposed the following quantum analogue of Fisher information
\begin{align*}
I_{\rm F}(\rho_\theta) = \frac{1}{4} \operatorname{tr}(\rho_\theta L_\theta^2),
\end{align*}
where $L_\theta$   is  the symmetric logarithmic derivative (SLD) of  $ \rho_\theta $ determined by
\begin{align*}
 \frac{\rm \partial}{{\rm \partial} \theta}\rho_\theta = \frac{1}{2} (L_\theta\rho_\theta +\rho_\theta L_\theta), \qquad \theta\in\mathbb R,
\end{align*}
which is a quantum analogue of Eq. (\ref{logd}).
%This quantify plays a crucial role in quantum parameter estimation.

Another natural approach to quantum Fisher information is to utilize Eq. (\ref{D2}) to  make the classical-to-quantum extension, and  the following quantum analogue of  classical Fisher information
\begin{align}
I_{\rm H}(\rho_{\theta})=4{\rm tr}\Big(\frac{\rm \partial}{{\rm \partial} \theta}{\sqrt{\rho_\theta}}\Big)^2 \label{QF2}
\end{align}
is obtained.

Let $H$  be the Hamiltonian of a quantum system, which encodes interactions driving quantum state evolution.
%a Hermitian operator on a Hilbert space $\mathcal H$, representing total energy with real eigenvalues for measurable energy levels.
The dynamics of a quantum state, described by the density matrix $\rho_{\theta}$, are
governed by the Liouville-von Neumann equation
\begin{align*}
i \frac{\rm \partial}{{\rm \partial} \theta}\rho_\theta =[H,\rho_\theta ], \qquad \theta\in\mathbb R,
\end{align*}
where $[X,Y]=XY-YX$ denotes commutator between operators  $X$ and $Y$.  The solution of this equation is the quantum state obtained after the evolution
\begin{align}\label{rhotheta}
\rho_\theta=e^{-i\theta H}\rho e^{i \theta H}.
\end{align}

For pure state  $\rho=|\psi\rangle\langle\psi|$, the Liouville-von Neumann equation reduces to the Schr\"odinger equation
\begin{align*}
i \frac{\rm \partial}{{\rm \partial} \theta}|\psi_{\theta}\rangle =  H |\psi_{\theta}\rangle, \qquad \theta\in\mathbb R,
\end{align*}
with solution $|\psi_{\theta}\rangle =e^{-i\theta H}|\psi\rangle$. This equation is fundamental and important in quantum information, enabling predictions of state evolution in quantum systems.

It has been proven that $ I_{\rm F}(\rho_\theta) $ is independent of $\theta$ when Eq. (\ref{rhotheta}) holds,  and thus simplifies to \cite{luo2004wigner}
\begin{align}\label{QFISLD}
I_{\rm F}(\rho, H) = \frac{1}{4} \operatorname{tr}(\rho L^2),
\end{align}
where $L$  satisfies
\begin{align*}
i [\rho, H]=\frac 12 (L\rho + \rho L).
\end{align*}
The quantum Fisher information $I_{\rm H}(\rho_{\theta})$ defined by Eq. (\ref{QF2}) is independent of the parameter $\theta$ if $\rho_{\theta}$ satisfies  Eq. (\ref{rhotheta}), and it holds that  \cite{luo2004wigner}
\begin{align*}
I_{\rm H}(\rho_{\theta}) = 8I_{\rm W}(\rho,H),
\end{align*}
where
\begin{align}\label{Wign}
I_{\rm W}(\rho,H)=-\frac{1}{2}{\rm tr}[\sqrt\rho,H]^2
\end{align}
is the celebrated Wigner-Yanase skew information \cite{Wign1963}.
%introduced by  Wigner and  Yanase in 1963  in  the context of quantum measurement theory \cite{Wign1963}.
%The QFI is significant in quantum metrology, where it sets the quantum Cram¨¦r-Rao bound, determining the ultimate precision limit for estimating parameters.
%={\rm tr}\rho H^2-{\rm tr}H\sqrt\rho H\sqrt \rho
Thus, the skew information can be regarded as a version of quantum Fisher  information. For pure states, it reduces to the variance. It is convex in state $\rho$, which is proven by Lieb in 1973 and instrumental in establishing  the strong subadditivity of von Neumann entropy \cite{Lieb1973}.

The relationship between $I_{\rm F}(\rho,H)$ and $I_{\rm W}(\rho,H)$ is given by
\begin{align}
I_{\rm W}(\rho,H)&\leq I_{\rm F}(\rho,H)\leq 2I_{\rm W}(\rho,H),
\end{align}
where the first inequality is saturated, i.e. $I_{\rm W}(\rho,H)= I_{\rm F}(\rho,H),$ if $\rho$ is pure or $[\rho,H]=0$ \cite{luo2004wigner}. Specifically, if
$[\rho,H]=0$,  both $I_{\rm W}(\rho,H)= I_{\rm F}(\rho,H)=0$, indicating no information about $\theta$.
%For pure states $\rho=|\psi\rangle\langle\psi|$, $I_{\rm W}(\rho,H)= I_{\rm F}(\rho,H)$.

For any mixed state $\rho$ with orthogonal spectral representation
\begin{align}\label{osp}
\rho=\sum_m \lambda_m |\phi_m\rangle\langle \phi_m|,
\end{align}
where $\{\lambda_m\}$ are the different  positive  eigenvalues of $\rho$ and $\{|\phi_m\rangle\}$ are the corresponding eigenstates forming an orthonormal base, the Wigner-Yanase skew information and  the quantum Fisher information  based on the symmetric logarithmic derivative   can be explicitly calculated as \cite{luo2004wigner}
\begin{align*}
I_{\rm W}(\rho,H)&=\frac12 \sum_{m,n} (\lambda_m^{\frac12}-\lambda_n^{\frac12})^2 |\langle \phi_m|H|\phi_n\rangle|^2,\\
I_{\rm F}(\rho,H)&=\frac12 \sum_{m,n} \frac{(\lambda_m-\lambda_n)^2 }{\lambda_m+\lambda_n}|\langle \phi_m|H|\phi_n\rangle|^2.
\end{align*}

In 2008, Hansen introduced the metric-adjusted skew information based on  operator monotone functions \cite{hansen2008metric,Petz1996}. It is  a generalized  quantum Fisher information that encompasses the well-known Wigner-Yanase skew information defined by Eq. (\ref{Wign}) and the quantum Fisher information based on the symmetric logarithmic derivative defined by Eq. (\ref{QFISLD}) as two important special cases. As a versatile and significant measure of information content, the metric-adjusted skew information has attracted significant interest and found extensive applications   in various fields, including uncertainty relations \cite{luo2005heisenberg,gibilisco2007uncertainty,yanagi2011metric,ren2021tighter}, correlations \cite{li2023probing,ren2024quantifying}, quantum interference \cite{gibilisco2021unified,fu2025complementarity}, coherence  \cite{sun2022quantifying}, and asymmetry \cite{takagi2019skew}.

\section{Quantifier of Complexity}
Inspired by the Fisher discord, which is the difference between the quantum Fisher information based on the symmetric logarithmic derivative and the Wigner-Yanase skew information, we define 
\begin{align}
C(\rho,H)=I_{\rm F}(\rho,H)-I_{\rm W}(\rho,H) \label{CM}
\end{align}
as a quantifier for  complexity of a quantum state $\rho$ relative to a Hamiltonian $H$. 

%Inspired by the Fisher discord, difference between quantum Fisher information in terms of symmetric logarithmic derivative and Wigner-Yanase skew information, let us define that for a quantum state $\rho,$
%\begin{align}
%C(\rho,H)=I_{\rm F}(\rho,H)-I_{\rm W}(\rho,H) \label{CM}
%\end{align}
%as  a quantifier for complexity of $\rho$ relative to the Hamiltonian $H.$

 {\bf Proposition 1}.  The quantifier  $C(\rho,H)$   for complexity has the following properties:

\noindent (1) $0\leq C(\rho,H)\leq I_{\rm W}(\rho,H).$

%(2) $C(\rho, H)=0$ if $\rho$ is pure or $[\rho, H]=0.$

\noindent (2) For mixed states $\rho$ with orthogonal  spectral representation defined by Eq.  (\ref{osp}), we have
\begin{align}\label{Crhoh}
C(\rho, H)= \sum_{m,n} \frac{\lambda_m^{\frac12}\lambda_n^{\frac12}(\lambda_m^{\frac12}-\lambda_n^{\frac12})^2 }{\lambda_m+\lambda_n}|\langle \phi_m|H|\phi_n\rangle|^2.
\end{align}

\noindent (3) For any unitary operator $U,$ it holds that
\begin{align*}
C(U \rho U^\dag,H)&=C(\rho,U^\dag H U) .
\end{align*}
In particular,
\begin{align*}
C(e^{i\theta H} \rho e^{-i\theta H},H)&=C(\rho, H) ,\qquad \theta\in[0,2\pi).
\end{align*}

\noindent (4) For any $\omega\in\mathbb R,$ it holds that
\begin{align*}
C(\rho ,H+\omega {\bf 1})&=C(\rho, H),\\
C(\rho ,\omega H)&=\omega^2C(\rho, H),
\end{align*}
where ${\bf 1}$ denotes the identity operator. Moreover,
\begin{align*}
C(\rho ,H_1+H_2)+C(\rho ,H_1-H_2)&= 2\sum_{j=1,2}C(\rho, H_j)
\end{align*}
for any two Hamiltonians  $H_1$  and $H_2$  of the  quantum system.

\noindent (5) $C(\rho,H)$ is additive under tensoring \cite{hansen2008metric}, in the sense that for any product states $\otimes_j \rho_j$ and composite Hamiltonian  operators $H=\sum_{j} \omega_j H_j\otimes {\bf 1}_{j^c},$ where $j^c:=\{i:\ i\neq j\}$ and $\omega_j\in\mathbb R,$ it holds that
\begin{align*}
C(\otimes_j \rho_j,H)&=\sum_j \omega_j^2 C(\rho_j, H_j).
\end{align*}
%which is direct from the tensor product  additivity of $I_{\rm F}$ and $I_{\rm W}.$
Particularly,
\begin{align}\label{additive}
\nonumber &\quad  C(\rho_1\otimes \rho_2,H_1\otimes{\bf 1}+{\bf 1}\otimes H_2)\\
&=C(\rho_1, H_1)+C(\rho_2, H_2),
\end{align}
for any quantum states $\rho_1$ and $\rho_2$, and any Hamiltonian operators $H_1$ and $H_2,$ on parties   $1$ and $2$, respectively.   In particular,
 \begin{align}
C(\rho_1\otimes \rho_2,H_1\otimes{\bf 1})=C(\rho_1, H_1).
\end{align}

\noindent (6) For any composite systems with subsystem Hamiltonian operators $H_1$ and $H_2,$ it holds that
\begin{align*}
&\quad C(\rho, H_1\otimes {\bf 1}+{\bf 1}\otimes H_2)+C(\rho, H_1\otimes {\bf 1}-{\bf 1}\otimes H_2)\\
&=2C(\rho, H_1\otimes {\bf 1})+2C(\rho,{\bf 1}\otimes H_2).
\end{align*}
%where ${\bf 1}$ stands for the identity operator of the subsystem.

\noindent (7) $C(\rho,H)$  is additive under direct sum in the sense that
\begin{align*}
C\Big(\bigoplus\limits_{i} p_i\rho_i,\bigoplus\limits_{i} H_i\Big)=\sum_i p_iC(\rho_i, H_i)
\end{align*}
for any quantum states $\rho_i$,  any  Hamiltonian operators $H_i$, and any probability distribution  $\{p_i\}$.
%This result is directly derived from the additivity of  quantum Fisher information and Wigner-Yanase skew information.

The above results follow readily  from the properties of the  quantum Fisher information  $I_{\rm F}(\rho,H)$ and the skew information  $I_{\rm W}(\rho,H)$ and direct verification.

{\bf Proposition 2.}
The complexity of a quantum state vanishes if the state is a pure state or a stable state (that is, a state that commutes with the Hamiltonian). In other words,  $C(\rho, H)=0$ if $\rho$ is pure or $[\rho, H]=0.$

%The complexity of the quantum state vanishes if it is a pure state or a stable state (that is, it commutes with the Hamiltonian), in other words, $C(\rho, H)=0$ if $\rho$ is pure or $[\rho, H]=0.$

We remark that pure states and stable states are merely sufficient but not necessary conditions for the disappearance of complexity in terms of Fisher discord. To see this, consider the mixture of the Bell state $|\psi_{0,1}\rangle=(|0\rangle+|1\rangle)/\sqrt{2}$  and pure state $|2\rangle$ in a single-mode Bosonic system, that is
\begin{align*}
\rho=\frac{1}{2}|\psi_{0,1}\rangle\langle \psi_{0,1}|+\frac{1}{2}|2\rangle\langle2|.
\end{align*}
We find that
\begin{align*}
C(\rho,a^\dag a)=0,
\end{align*}
while
$[\rho,a^\dag a]\neq0.$
The Hamiltonian $a^\dag a$ considered here represents the number operator in quantum optics.

%We remark that  pure states and stable states are merely sufficient but not necessary conditions for the disappearance of complexity in terms of Fisher discord. To see this, consider the mixture of the Bell state $|\psi_{0,1}\rangle=(|0\rangle+|1\rangle)/\sqrt{2}$  and pure state $|2\rangle$ in a single-mode Bosonic system, that is
%\begin{align*}
%\rho=\frac{1}{2}|\psi_{0,1}\rangle\langle \psi_{0,1}|+\frac{1}{2}|2\rangle\langle2|,
%\end{align*}
%we have
%\begin{align*}
%C(\rho,a^\dag a)=0,
%\end{align*}
%while
%$[\rho,a^\dag a]\neq0.$
%The Hamiltonian $a^\dag a$ considered here represents the number operator in quantum optics.
%with $[a,a^\dagger]={\bf 1}$ satisfying the canonical commutation relation.

 \section{Illustrations}

In quantum mechanics,  the  Hamiltonian operator  of a system  represents the  total energy of that system. Hamiltonian is of fundamental importance in quantum theory from its intimate link to the energy spectrum and the time evolution of quantum states via the   Schr\"odinger equation \cite{luo2004fast,vogel2006quantum,berezin2012schrodinger}. Next, we illustrate the complexity quantifier $C(\rho,H)$  by evaluating it
for several paradigmatic Hamiltonians  in both discrete-variable systems and continuous-variable quantum systems, revealing quantitative insights of the quantum states from the perspective of their dynamical evolution.

\subsection{Complexity of discrete-variable systems}
In this subsection, we illustrate the complexity quantifier
 through  a general  qubit state and Hamiltonian, and reveal
some quantitative features of complexity   from the perspective
 of quantum dynamic.

Any qubit state can be represented in the basis $\{|0\rangle,|1\rangle \}$ as
\begin{align}
\nonumber \rho  =\frac12\Big( {\bf 1}+\sum_{j=1}^3 r_j \sigma_j\Big)=\frac12
\Bigg(
\begin{matrix}
1+r_3 & r_1-ir_2\\
r_1+ir_2 & 1-r_3
\end{matrix}
\Bigg),
\nonumber
\end{align}
where $r=\sqrt{\sum_{j=1}^3 r_j^2}\leq1, r_j\in\mathbb R,$ and $\sigma_j$ is the Pauli matrices for any $j=1,2,3$.
The  spectral decomposition of $\rho $ is
\begin{align*}
\rho  =\lambda_1 |\phi_1\rangle\langle \phi_1|+\lambda_2  |\phi_2\rangle\langle \phi_2|,
\end{align*}
with the eigenvectors
\begin{align*}
|\phi_1\rangle &=\frac{1}{\sqrt{2r(r-r_3)}}\Big((r_1-ir_2) |0\rangle-(r_3-r)|1\rangle \Big),\\
|\phi_2\rangle &=\frac{1}{\sqrt{2r(r+r_3)}}\Big((r_1-ir_2) |0\rangle-(r_3+r)|1\rangle \Big),
\end{align*}
and the corresponding  eigenvalues
\begin{align*}
\lambda_1=\frac{1}{2}(1+r),\qquad  \lambda_2=\frac{1}{2}(1-r).
\end{align*}
Let $H$ be a Hamiltonian for a two-level quantum system.  By Eq.  (\ref{Crhoh}), we have
\begin{align*}
C(\rho ,H)= \frac 1 2 \big(|r|^2-1+\sqrt{1-|r|^2}\big)|\langle\phi_1|H|\phi_2\rangle|^2.
\end{align*}
If $\langle\phi_1|H|\phi_2\rangle\neq0$ which indicates $[\rho,H]\neq0$ for any two dimensional system, then  $C(\rho ,H)=0$ if and only if  $\rho $ is a pure state, i.e., $r=1$.

The above results lead to the following proposition.

{\bf Proposition 3.}
The complexity of a qubit  state vanishes if and only if  it is a pure state or a stable state (that is, it commutes with the Hamiltonian).  In other words, $C(\rho , H)=0$ if  and only if $\rho $ is pure or $[\rho , H]=0.$

\subsection{Complexity of continuous-variable systems}

A single-mode Bosonic  field, fundamental in quantum optics, describes a quantized harmonic oscillator, such as a single electromagnetic field mode. It is
characterized by the annihilation operator $a$  and creation operator  $a^{\dag}$ satisfying
the commutation relation
\begin{align*}
[a, a^{\dag}]=\bf{1}.
\end{align*}
To further illustrate the complexity quantifier based on Fisher discord, we evaluate the complexity of
 various Bosonic states relative to several prototypical Hamiltonians.

\vskip 0.2cm

1. {\sl Complexity relative to   $a^\dag a$}

First consider the Hamiltonian $H=a^\dag a$. In
quantum optics, $a^\dag a$ represents the number operator, whose eigenstates correspond to the Fock states, i.e., $a^{\dag}a|n\rangle=n|n\rangle, \ n=0,1\cdots $. This operator is fundamental for several reasons: (i) It governs the free evolution of a harmonic oscillator, describing the energy of the field mode (up to a constant). (ii) Its eigenstates, the Fock states $|n\rangle, \ n=0,1\cdots$, form a standard orthogonal basis for the Hilbert space and are critical for analyzing photon statistics. (iii) It underpins the definition of coherent and squeezed states, serving as a benchmark for classicality versus nonclassicality. (iv) It plays a central role in characterizing quantum optical phenomena, such as sub-Poissonian statistics and photon bunching. By evaluating  $C(\rho,a^\dag a)$, we gain
insights into the complexity of the state $\rho$ in terms of its photon number distribution and nonclassical features under number-conserving dynamics.

Apparently, for the  rotation operators $U=e^{i\omega a^\dag a}, \omega\in[0,2\pi),$ we have
\begin{align*}
C( U \rho U^\dag,a^\dag a)&=C(\rho,a^\dag a).
\end{align*}
However, $C(\rho, a^\dag a)$ is not displacement invariant, except for the states defined by Eq. (\ref{osp}) satisfying $\lambda_m\lambda_{m+1}=0$ for any $m.$

To further illustrate the complexity  quantifier relative to the Hamiltonian  $a^\dag a$,  we evaluate the complexity of  several prototypical Bosonic  states.

{\bf Proposition 4.}  The  complexity relative to Hamiltonian  $a^\dag a$ of  Fock-diagonal states,  displaced Fock-diagonal states  and squeezed Fock-diagonal states are as follows.

\noindent (1)  For any Fock-diagonal state
\begin{align}\label{FD}
\rho_{\rm D}=\sum_{n=0}^\infty \lambda_n |n\rangle\langle n|,
\end{align}
where $ \lambda_n\geq0$ for each $n$ with $ \sum_n\lambda_n=1$, the amount of complexity is
\begin{align*}
C(\rho_{\rm D},a^\dag a)=0.
\end{align*}
%which  is  directly from that $[\rho_{\rm D},a^\dag a]=0$.

\noindent  (2)  For any  displaced Fock-diagonal state $D_z \rho_{\rm D} D_z^\dag$, we have
\begin{align*}
C(D_z\rho_{\rm D} D_z^\dag,a^\dag a)&=|z|^2C(\rho_{\rm D},X_{\theta}), \qquad \theta \in [0,2\pi).
\end{align*}
Here, $ D_z=e^{z a^\dag-z^* a},\ z\in\mathbb C$  are  displacement operators,  and  $X_\theta=e^{-i\theta}a+e^{i\theta} a^\dag, \ \theta \in [0,2\pi)$.

\noindent (3) For any squeezed diagonal state  $S_\zeta\rho_{\rm D} S_\zeta^\dag,$  we have
\begin{align*}
&\quad C(S_\zeta\rho_{\rm D} S_\zeta^\dag,a^\dag a)\\
&=\frac{\sinh^2 (2|\zeta|)}4 C(\rho_{\rm D},a^{\dag 2}+a^2)\\
&=\frac{\sinh^2(2|\zeta|)}{2}\sum_{n=0}^\infty\frac{\lambda_n^{\frac12} \lambda_{n+2}^{\frac12} (\lambda_{n+2}^{\frac12} -\lambda_{n}^{\frac12} )^2}{\lambda_{n+2}+\lambda_n}(n^2+3n+2)\\
&=\frac{\sinh^2(2|\zeta|)}{4}C(\rho_{\rm D},\Lambda_\theta),
\end{align*}
for any $\theta \in [0,2\pi)$. Here $S_\zeta=e^{(\zeta^* a^2-\zeta a^{\dag 2})/2}, \ \zeta=|\zeta|e^{i\arg \zeta}\in\mathbb C$ are squeezing operators, and $\Lambda_{\theta}=e^{-i\theta}a^2+e^{i\theta} a^{\dag2},\ \theta \in [0,2\pi)$.

The above results can be directly  verified.
Next, we specify  to the  thermal  state 
\begin{align}\label{thermal}
\tau_\lambda=(1-\lambda)\sum_{n=0}^\infty \lambda^n |n\rangle\langle n|,\qquad  \lambda\in[0,1)
\end{align}
as an important example of Fock-diagonal states.

For any  displaced thermal states $D_z \tau_\lambda D_z^\dag,$  we have
\begin{align*}
C(D_z \tau_\lambda D_z^\dag,a^\dag a)=\frac{2|z|^2\sqrt{\lambda}(1-\sqrt\lambda)^2}{1-\lambda^2},
\end{align*}
which monotonically increases in  the displacement parameter $|z|$ for fixed $\lambda,$ increases first and then decreases in  the noise parameter $\lambda$ for fixed $z,$ and reaches its maximum value at
$\lambda=\lambda_0,$ where
\begin{align}\label{lambda0}
\lambda_0=:2+\sqrt{5}-2\sqrt{2+\sqrt{5}}\approx0.1197.
\end{align}

For any squeezed thermal states $S_\zeta\tau_\lambda S_\zeta^\dag,$ we have
\begin{align*}
 C(S_\zeta\tau_\lambda S_\zeta^\dag,a^\dag a)=\frac{\lambda \sinh^2(2|\zeta|) }{1+\lambda^2 },
\end{align*}
which is monotonically increasing of the squeezing strength $|\zeta|$ for fixed $\lambda,$ and of the noise parameter $\lambda$ for fixed $\zeta.$
In particular, for thermal states ($\zeta=0,$ Fock-diagonal as mentioned above), we have $C(\tau_\lambda,a^\dag a)=0.$

For the Gaussian states $\rho_g=D_z S_\zeta\tau_\lambda  S^{\dag}_\zeta D^\dag_z$, we have
\begin{align*}
C(\rho_g,a^\dag a)=\frac{\lambda\sinh^2(2|\zeta|)}{1+\lambda^2}+\frac{2\sqrt{\lambda}(1-\sqrt\lambda)^2|\beta_{z,\zeta}|^2}{1-\lambda^2},
\end{align*}
where $\beta_{z,\zeta}=z\cosh{|\zeta|}-z^*e^{i\arg\zeta}\sinh|\zeta|$. 
For fixed $\lambda$ and $\zeta,$   $C(\rho_g,a^\dag a)$ monotonically increases in  the displacement parameter $|z|.$
% and  then $$|\beta_{z,\zeta}|^2=|z|^2\Big(\cosh(2|\zeta|)-\sinh{(2|\zeta|)}\cos{(2\arg z-\arg\zeta)}\Big).$$
%\begin{align*}
%C(\rho_g,a^\dag a)=\frac{\lambda\sinh^2(2|\zeta|)}{1+\lambda^2}+\frac{2\sqrt{\lambda}(1-\sqrt\lambda)^2|z|^2}{1-\lambda^2}\Big(\cosh(2|\zeta|)-\sinh{(2|\zeta|)}\cos{\theta}\Big),
%\end{align*}
For fixed $\lambda$, $|\zeta|$ and  $|z|$, $C(\rho_g,a^\dag a)$  monotonically increases in  the  parameter $(2\arg z-\arg\zeta)$ in  $[0,\pi)$, and monotonically decreases in $[\pi,2\pi].$ For fixed $\lambda$,  $(2\arg z-\arg\zeta)$ and  $|z|,$ $C(\rho_g,a^\dag a)$  monotonically increases in  the  squeezing  parameter $|\zeta|$ if $\cos{(2\arg z-\arg\zeta)}\leq0$, and monotonically decreases first and then increases  in  the  squeezing parameter $|\zeta|$ if $\cos{(2\arg z-\arg\zeta)}>0.$

\begin{figure}[htbp]
 \centering
{
  \includegraphics[width=0.48\textwidth]{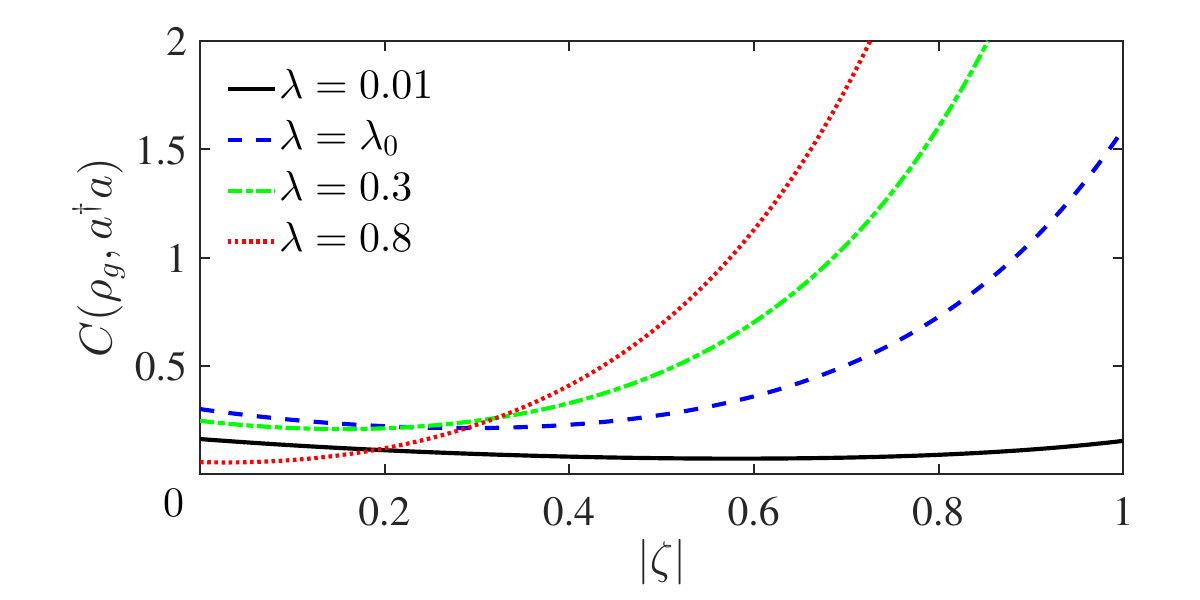}
}
 \caption{Complexity $C(\rho_{g},a^\dag a)$ of Gaussian state $\rho_{g}$ relative  to  Hamiltonian $a^\dag a$ versus the squeezing parameter $|\zeta|$ for $\lambda=0.01$ (black solid  line), $\lambda_0$ (blue dashed line, defined by Eq. (\ref{lambda0})),  $0.3$ (green dash-dotted   line), and $0.8$  (red dotted  line)  with fixed  $|z|=1$  and  $2\arg z-\arg\zeta=0$.}\label{fig1}
\end{figure}

The graphs of $C(\rho_g,a^\dag a)$  versus $|\zeta|$  with fixed  $|z|=1$,  $2\arg z-\arg\zeta=0$ and $\lambda=0.01,  \ \lambda_0,\ 0.3, \ 0.8$ are presented in Fig.  \ref{fig1}, with $\lambda_0$ defined by Eq. (\ref{lambda0}), which reveal that  $C(\rho_g,a^\dag a)$  initially  decreases monotonically and then increases in  $|\zeta|.$
However, the minimum point and its corresponding value vary from the parameters  $\lambda$ and  $|z|$.
\begin{figure}[h]
 \centering
{
\includegraphics[width=0.48\textwidth]{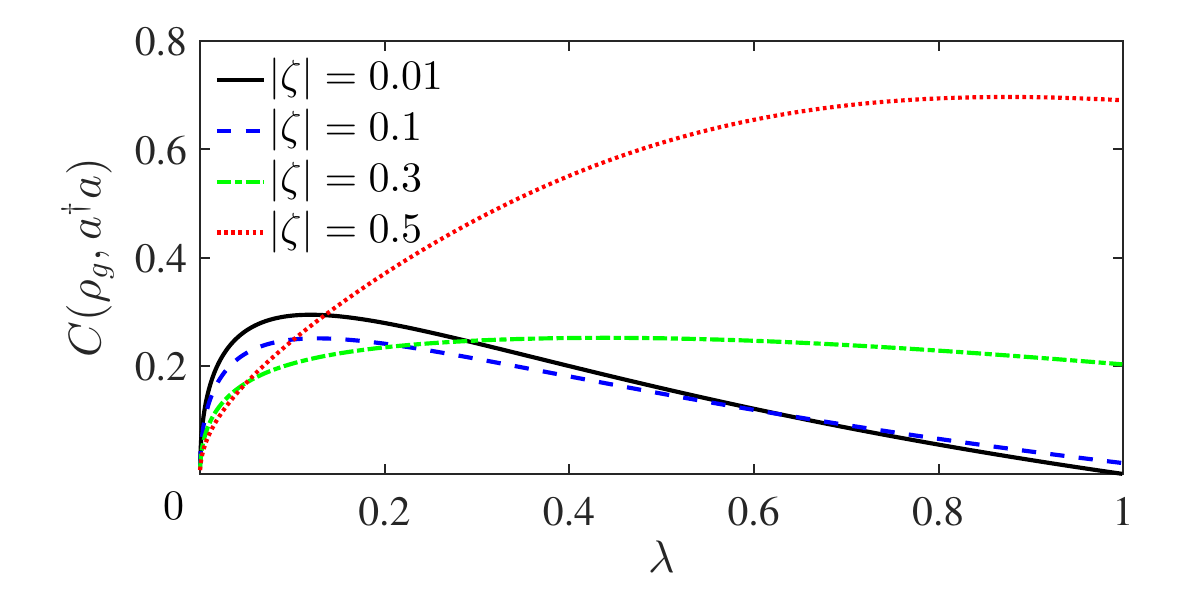}
}
 \caption{Complexity $C(\rho_{g},a^\dag a)$ of Gaussian state $\rho_{g}$ relative  to  Hamiltonian $a^\dag a$ versus the noise parameter $\lambda$ for $|\zeta|=0.01$ (black solid line), $0.1$ (blue dashed line), $0.3$ (green dash-dotted   line),  $0.5$ (red dotted  line)   with fixed  $|z|=1$  and  $2\arg z-\arg\zeta=0$.}\label{fig2}
\end{figure}

For fixed $z$ and $\zeta,$
$C(\rho_g,a^\dag a)$ monotonically increases in  the noise parameter $\lambda$ in $(0,\lambda_0].$
When $\lambda>\lambda_0$, the monotonicity of $C(\rho_g,a^\dag a)$  in  parameter $\lambda$  depends on the specific parameter values of  $z$ and $\zeta$. The graphs of $C(\rho_g,a^\dag a)$  versus $\lambda$  with fixed  $|z|=1$, $2\arg z-\arg\zeta=0$ and $|\zeta|=0.01, \ 0.1,\   0.3, \ 0.5$ are shown in Fig. \ref{fig2}, which show that  $C(\rho_g,a^\dag a)$  initially  increases monotonically and then decreases in  $\lambda.$ However, the maximum point and its corresponding value vary depending on the parameters  $\zeta$  and  $z$.

%untitiled34-figure2
%\begin{align*}
%C(\rho_g,a^\dag a)=\frac{\lambda\sinh^2(\frac{\pi}{2})}{1+\lambda^2}+\frac{2\sqrt{\lambda}(1-\sqrt\lambda)^2}{1-\lambda^2}\Big(\cosh(\frac{\pi}{2})-\sinh{(\frac{\pi}{2})}\cos{(2\arg z-\arg\zeta)}\Big),
%\end{align*}

\begin{figure}[htbp]
 \centering
{
  \includegraphics[width=0.48\textwidth]{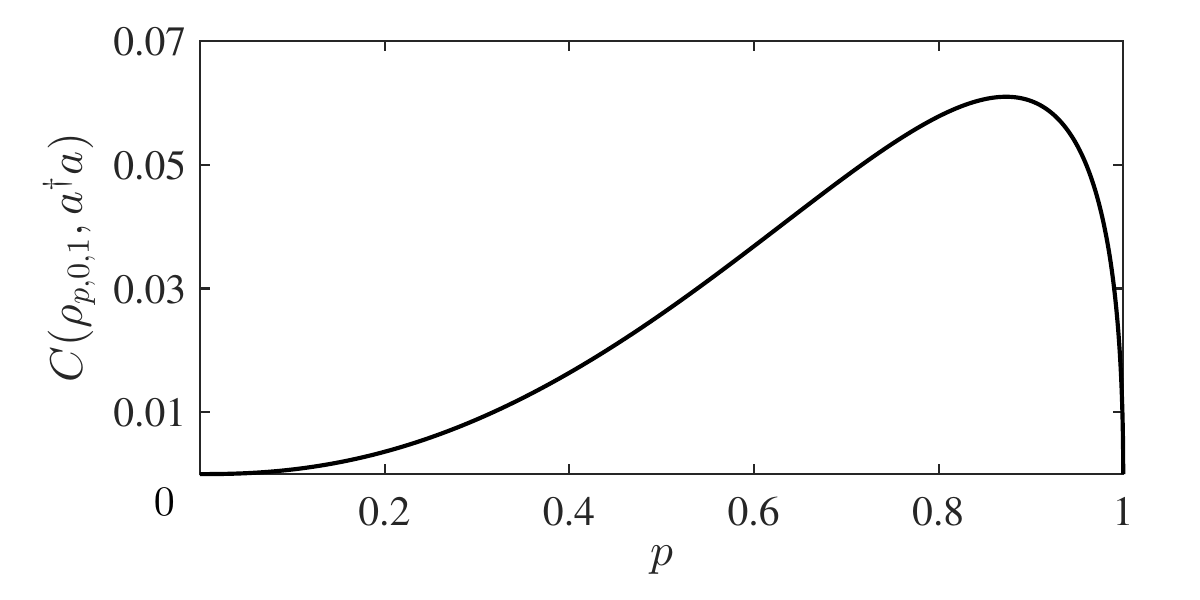}
}
 \caption{Complexity $C(\rho_{p,0,1},a^\dag a)$ of the  mixture state $\rho_{p,0,1}$ relative  to  Hamiltonian $a^\dag a$ versus the parameter $p\in[0,1]$.
 }\label{fig3}
\end{figure}

For the mixture of the pure states $|\psi_{m,n}\rangle=(|m\rangle+|n\rangle)/\sqrt{2}, \ m, n\in\mathbb{Z}, \ m\neq n$ and vacuum state $|0\rangle,$ that is
\begin{align}\label{rhopmn}
\rho_{p,m,n}=p|\psi_{m,n}\rangle\langle \psi_{m,n}|+(1-p)|0\rangle\langle0|
\end{align}
with $p\in(0,1),$ it holds that when $mn\neq0$,
\begin{align*}
C(\rho_{p,m,n},a^\dag a)=0,
\end{align*}
and when $mn=0$ ($n>0$ without loss of generality),
\begin{align*}
C(\rho_{p,0,n},a^\dag a)=\frac{n^2p^2f_{p}}{4\big((1-p)^2+p^2\big)},
\end{align*}
where
\begin{align}\label{f}
f_{p}=\sqrt{2p(1-p)}\big(1-\sqrt{2p(1-p)}\big).
\end{align}
$C(\rho_{p,0,n},a^\dag a)$  vanishes if $p=0$ or $1$ because the state $\rho_{p,0,n}$ degenerates into a Fock-diagonal state, monotonically increases in  $n$ for fixed $p,$ increases first and then decreases related to $p$ for fixed $n,$ and reaches its maximum value at $p\approx0.8731.$  The complexity quantity $C(\rho_{p,0,1},a^\dag a)$ of $\rho_{p,0,1}$ in  $p$ is shown by Fig. \ref{fig3}.
%$1/24 (18 + \sqrt{6 (-12 + (33 (9 - 4 \sqrt3))^{1/3} + (33 (9 + 4 \sqrt3))^{1/3})} + \sqrt{-6 (33 (9 - 4 \sqrt3))^{1/3} -6 (33 (9 + 4 \sqrt3))^{1/3} + 36 (-4 + \sqrt{6/(-12 + (33 (9 - 4 \sqrt3))^{1/3} + (33 (9 + 4 \sqrt3))^{1/3})})}$
In particular, when $p=1/2$,
\begin{align}\label{rho1/20n}
C(\rho_{1/2,0,n},a^\dag a)=\frac{\sqrt{2}-1}{16}n^2,\qquad n>0.
\end{align}

%untitled6-figure3

%For the mixture\\
%$\qquad  \rho_{p,|\psi_{0,n}\rangle}=p|\psi_{0,n}\rangle\langle \psi_{0,n}|+(1-p)|0\rangle\langle0|,\quad p\in(0,1),$\\
%of the pure states $|\psi_{0,n}\rangle=(|0\rangle+|n\rangle)/\sqrt{2},\quad (n>0),$ and the vacuum state $|0\rangle$, we have $Q(\rho_{1/2,0,n},a^\dag a)=(3-2\sqrt{2})n^2/16$

\vskip 0.2cm

2. {\sl Complexity relative to  $e^{-i\theta}a+e^{i\theta} a^\dag$}

%\subsection{$H=X_\theta=e^{-i\theta}a+e^{i\theta} a^\dag$}
%=\sqrt{2}(X\cos{\theta}+P\sin{\theta})

Consider the complexity quantifier relative  to the fixed Hamiltonian $X_\theta=e^{-i\theta}a+e^{i\theta} a^\dag,\ \theta\in[0,2\pi)$, which represents a rotated quadrature operator, generalizing
the position and momentum observables defined as $X=(a+a^{\dag})/\sqrt{2}=X_0/\sqrt{2}$ and $P=(a-a^{\dag})/i\sqrt{2}=X_{\pi/2}/\sqrt{2}.$ Hamiltonian $X_\theta$ is fundamental in quantum optics, corresponding to the homodyne rotated quadrature, and enabling measurements of field quadratures at any phase $\theta$. This is crucial in homodyne detection, where the phase $\theta$ is tuned to probe position-like or momentum-like properties of the  field. By evaluating $C(\rho,X_\theta)$, we can gain quantitative insights into the complexity of Bosonic  state under quadrature-dependent dynamics.

% $ X_\theta=e^{-i\theta}a+e^{i\theta} a^\dag,\theta\in[0,2\pi)$ is the generalized position operator

The quantifier of complexity  $C(\rho,X_\theta)$  relative to $X_\theta$  is invariant with respect to any  displacement  operator $D_z$ in the sense that
\begin{align*}
C( D_z \rho D_z^\dag,X_\theta)=C(\rho,X_\theta),\qquad z\in\mathbb C,
\end{align*}
  which follows readily from 
\begin{align*}
C(  D_z\rho  D_z^\dag,X_\theta)&=C(\rho,  D_z X_\theta  D_z^\dag)\\
&= C(\rho, X_\theta +2{\rm Re}( e^{-i\theta}z){\bf 1})\\
&=C( \rho ,X_\theta).
\end{align*}
%$X=(a+a^{\dag})/\sqrt{2}$ and $P=(a-a^{\dag})/i\sqrt{2}$ are the position and momentum observables, respectively.
In contrast, $C(\rho,X_\theta)$ is not rotation invariant, as
\begin{align*}
C( e^{i\omega a^\dag a} \rho e^{-i\omega a^\dag a}, X_\theta)&=C(\rho,X_{\theta+\omega}),
\end{align*}
except for the displaced Fock-diagonal states.

In order to gain an intuitive understanding of  complexity, we evaluate   $C(\rho ,X_\theta)$ for various Bosonic   states.

{\bf Proposition 5.}  

\noindent (1)  For the  Fock-diagonal state $\rho_{\rm D}$ defined by Eq. (\ref{FD}), we have
\begin{align*}
C(\rho_{\rm D},X_\theta)=2 \sum_{n=0}^\infty \frac{ \lambda_{n}^{1/2}\lambda_{n+1}^{1/2}(\lambda_{n}^{1/2}-\lambda_{n+1}^{1/2})^2 }{\lambda_n+\lambda_{n+1}} (n+1),
\end{align*}
which is independent of parameter $\theta$.

\noindent  (2) For the squeezed diagonal state  $S_\zeta\rho_{\rm D} S_\zeta^\dag,$ we have
 \begin{align*}
C(S_\zeta\rho_{\rm D} S_\zeta^\dag,X_\theta)&=|\beta_{\theta,\zeta}|^2 C(\rho_{\rm D},X_\theta),
\end{align*}
where
 \begin{align}\label{betatheta}
\beta_{\theta,\zeta}=e^{i\theta}\cosh |\zeta|-e^{-i(\theta-\arg \zeta)}\sinh |\zeta|.
\end{align}

Take some typical diagonal states and  squeezed diagonal states for example.

\begin{figure}[htbp]
 \centering
{
 \includegraphics[width=0.48\textwidth]{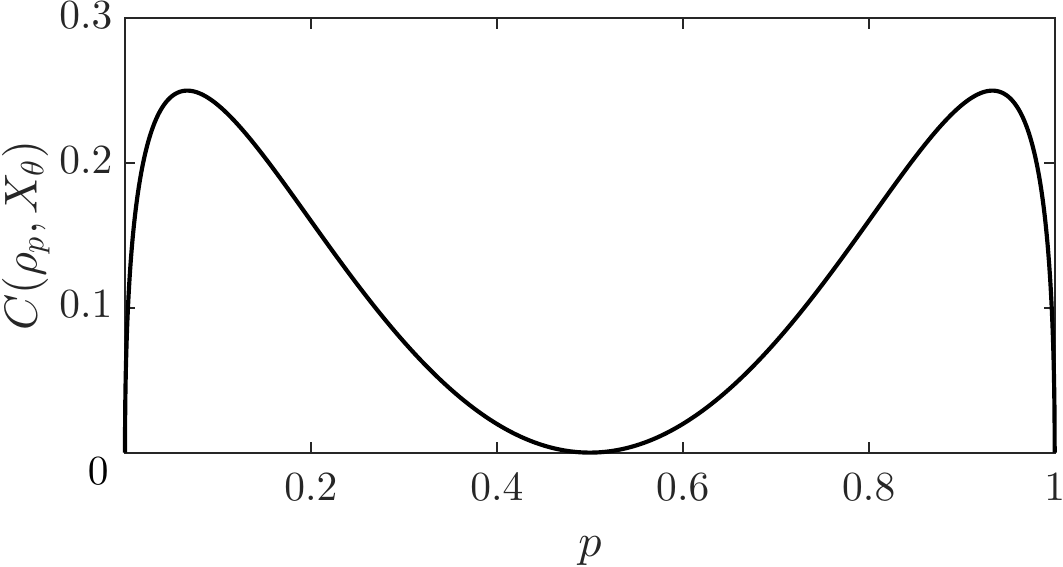}
}
 \caption{Complexity $C(\rho_p,X_\theta)=g_p$ of the state $\rho_{p}$ relative  to  Hamiltonian $X_\theta$ versus the parameter  $p\in[0,1]$.
 }\label{fig4}
\end{figure}

For $\rho_p=p|0\rangle\langle0|+(1-p)|1\rangle\langle1|,$ we have
\begin{align}\label{rhop}
C(\rho_p,X_\theta)&=2\sqrt{p(1-p)}\big(\sqrt{p}-\sqrt{1-p}\big)^2=:g_p,
\end{align}
which is shown in Fig. \ref{fig4}.
Apparently, it is neither convex nor concave in $p$. Moreover, it is not monotonic in $p,$ reaches its minimum value 0 at $p=0,1/2,1,$  and reaches its maximum value $1/4$ at $p=(2\pm\sqrt3)/4.$

In contrast, for $\rho_{p,k}=p|0\rangle\langle0|+(1-p)|k\rangle\langle k|,\ k\geq2,$  we have
\begin{align*}
C(\rho_{p,k},X_\theta)&=0.
\end{align*}
In general,
\begin{align*}
C(\rho_{p,k},e^{-i\theta}a^{l}+e^{i\theta} a^{\dag l})=g_p k!\delta_{k,l},
\end{align*}
where $g_p$ is defined by Eq. (\ref{rhop}), and $\delta_{k,l}=1$  if $k=l,$ otherwise, $\delta_{k,l}=0$.
%\begin{align*}
%C(\rho_{p,k},e^{-i\theta}a^{l}+e^{i\theta} a^{\dag l})=2\sqrt{p(1-p)}\big(1-2\sqrt{p(1-p)}\big)k!\delta_{k,l}.
%\end{align*}
This indicates that the complexity of the quantum state $\rho_{p,k}$ can be detected by the Hamiltonian $e^{-i\theta}a^{l}+e^{i\theta} a^{\dag l}$ with $l=k$.

For any thermal states $\tau_\lambda$ defined by Eq. (\ref{thermal}), we have
\begin{align*}
C(\tau_\lambda,X_\theta)=\frac{2\sqrt\lambda(1-\sqrt\lambda)^2}{1-\lambda^2},
\end{align*}
which is monotonically increasing first and then decreasing in  the noise parameter $\lambda$,
and  reaches its maximum value  0.3003 when $\lambda=\lambda_0$ defined by Eq. (\ref{lambda0}).
%0.300283

For any  squeezed thermal states $S_\zeta\tau_\lambda S_\zeta^\dag,$  we have
\begin{align*}
C(S_\zeta\tau_\lambda S_\zeta^\dag,X_\theta)&= \frac{2\sqrt\lambda(1-\sqrt{\lambda})^2 |\beta_{\theta,\zeta}|^2}{1-\lambda^2},
\end{align*}
where $\beta_{\theta,\zeta}$ is defined by Eq. (\ref{betatheta}).

For the truncated thermal state  (no vacuum component $|0\rangle\langle 0|$)
\begin{align*}
\tau^{\rm t}_\lambda=\frac{1-\lambda}{\lambda}\sum_{n=1}^\infty \lambda^n  |n\rangle\langle n|,\qquad \lambda\in(0,1),
\end{align*}
we have
\begin{align*}
C(\tau^{\rm t}_\lambda,X_\theta)&=\frac{2\sqrt\lambda(1-\sqrt{\lambda})^2(2-\lambda)}{1-\lambda^2},
\end{align*}
which reaches its maximum value  0.5675 when $\lambda\approx0.1014$.

\begin{figure}[htbp]
 \centering
{
 \includegraphics[width=0.44\textwidth]{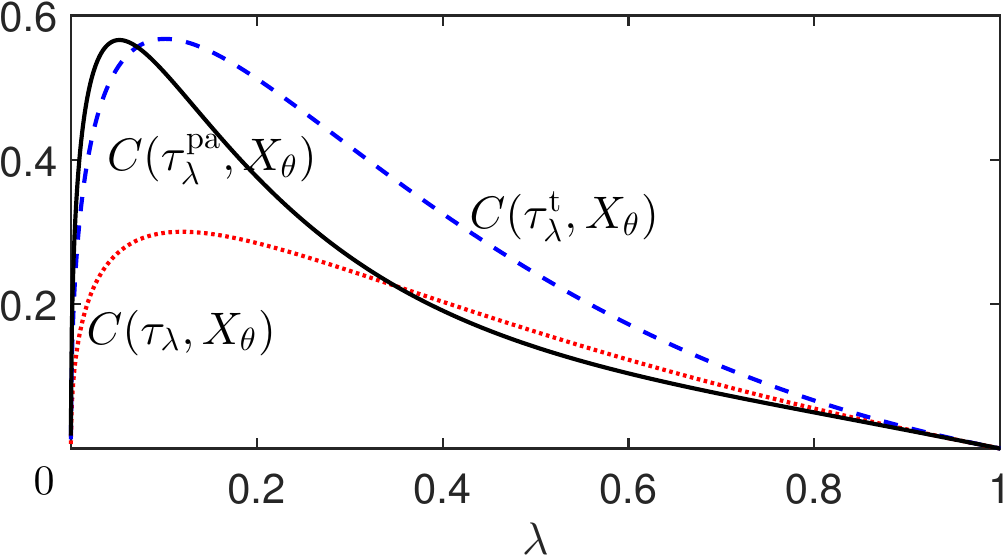}
}
 \caption{Complexity $C(\tau_\lambda,X_\theta)$ for thermal  state  $\tau_\lambda$ (red dotted line), $C(\tau^{\rm t}_\lambda,X_\theta)$ for truncated thermal state $\tau^{\rm t}_\lambda$ (blue dashed line)  and  $C(\tau^{\rm pa}_\lambda,X_\theta)$ for photon-added thermal state $\tau^{\rm pa}_\lambda$ (black solid line) versus noise parameter $\lambda\in(0,1)$.}\label{fig5}
\end{figure}

For the photon-added thermal state (no vacuum component)
\begin{align*}
\tau^{\rm pa}_\lambda=\frac{a^\dag\tau_\lambda a}{{\rm tr}a^\dag\tau_\lambda a}=\frac{(1-\lambda)^2}{\lambda}\sum_{n=1}^\infty \lambda^n  n|n\rangle\langle n|,\qquad \lambda\in(0,1),
\end{align*}
we have
\begin{align*}
C(\tau^{\rm pa}_\lambda,X_\theta)=\frac{2(1-\lambda)^2}{\sqrt\lambda}\sum_{n=1}^\infty p_{\lambda,n}\lambda^{n},
\end{align*}
where $$p_{\lambda,n}=\sqrt{n(n+1)^3}(\sqrt{(n+1)\lambda}-\sqrt{n})^2/\big((n+1)\lambda+n\big).$$
The comparison between the complexity $C(\tau_\lambda,X_\theta)$,  $C(\tau^{\rm t}_\lambda,X_\theta)$  and  $C(\tau^{\rm pa}_\lambda,X_\theta)$   as functions of  $\lambda$ is  shown in Fig. \ref{fig5}. All three of them  are  monotonically increasing first and then monotonically decreasing in  the  noise parameter $\lambda$.

%figure5-untitled

\begin{figure}[htbp]
 \centering
{
 \includegraphics[width=0.48\textwidth]{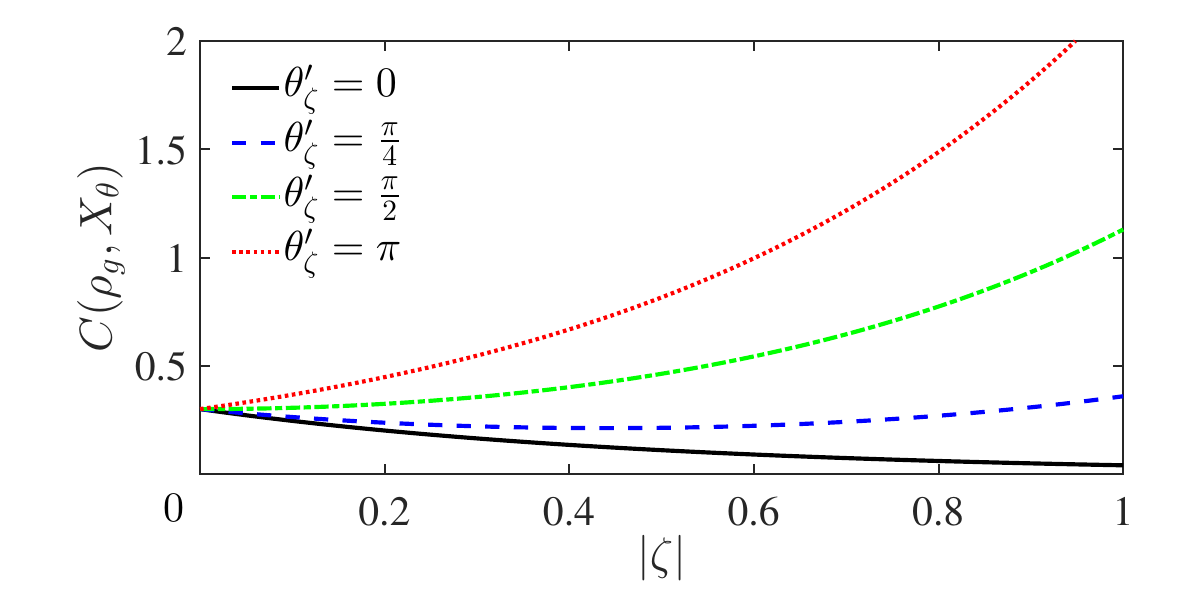}
}
\caption{Complexity $C(\rho_{g},X_\theta)$ of Gaussian state $\rho_{g}$ relative  to  Hamiltonian $X_\theta$ versus the squeezing parameter   $|\zeta|$ for $\theta^{\prime}_{\zeta}=0$ (black solid line), $\pi/4$ (blue dashed line), $\pi/2$ (green dash-dotted line), $\pi$ (red dotted line)  with fixed  $\lambda=\lambda_0$. Here, $\theta^{\prime}_{\zeta}=2\theta-\arg\zeta$. }\label{fig6}
\end{figure}

For the  Gaussian states
$\rho_g=D_z S_\zeta\tau_\lambda  S^{\dag}_\zeta D^\dag_z$, we have
\begin{align*}
C(\rho_g,X_\theta)&=C(S_\zeta\tau_\lambda S_\zeta^\dag,X_\theta)= \frac{2\sqrt\lambda(1-\sqrt{\lambda})^2 |\beta_{\theta,\zeta}|^2}{1-\lambda^2},
\end{align*}
where $\beta_{\theta,\zeta}$ is defined by Eq. (\ref{betatheta}), which is independent of the displacement parameter $z$. For fixed $\theta$ and $\zeta$, $C(\rho_g,X_\theta)$  initially increases monotonically and then decreases in  the noise parameter $\lambda$, reaching a maximum value  about  $0.3|\beta_{\theta,\zeta}|^2$ at $\lambda=\lambda_0$. For fixed $\lambda$  and $|\zeta|,$ $C(\rho_g,X_\theta)$  monotonically increases in  the  parameter $\theta^{\prime}_{\zeta}:=2\theta-\arg\zeta$ in $[0,\pi)$, and  decreases in $[\pi,2\pi].$ For fixed $\lambda$, $\arg\zeta$ and $\theta,$ $C(\rho_g,X_\theta)$  monotonically increases in  the  squeezing  parameter $|\zeta|$ when $\cos{\theta^{\prime}_{\zeta}}\leq0$ as shown by the green dash-dotted line and  red dotted line in Fig.  \ref{fig6}, and  monotonically decreases when $\cos\theta^{\prime}_{\zeta}=1$ as shown by the black solid  line  in Fig.  \ref{fig6}. However for fixed $\lambda$ and  $2\theta-\arg\zeta,$ $C(\rho_g,X_\theta)$ initially decreases monotonically and then increases in   $|\zeta|$ when $\cos\theta^{\prime}_{\zeta}\in(0,1)$, reaching its minimum value at $|\zeta|=\frac{1}{4}\ln\frac{1+\cos\theta^{\prime}_{\zeta}}{1-\cos\theta^{\prime}_{\zeta}},$ as exemplified by the blue dashed line in  Fig.  \ref{fig6} with $ \theta^{\prime}_{\zeta}=\pi/4$,  attaining   its minimum value at $|\zeta|=\frac{1}{2}\ln{(1+\sqrt{2})}\approx 0.4407.$
%untitled32-figure6

\begin{figure}[htbp]
 \centering
{
 \includegraphics[width=0.48\textwidth]{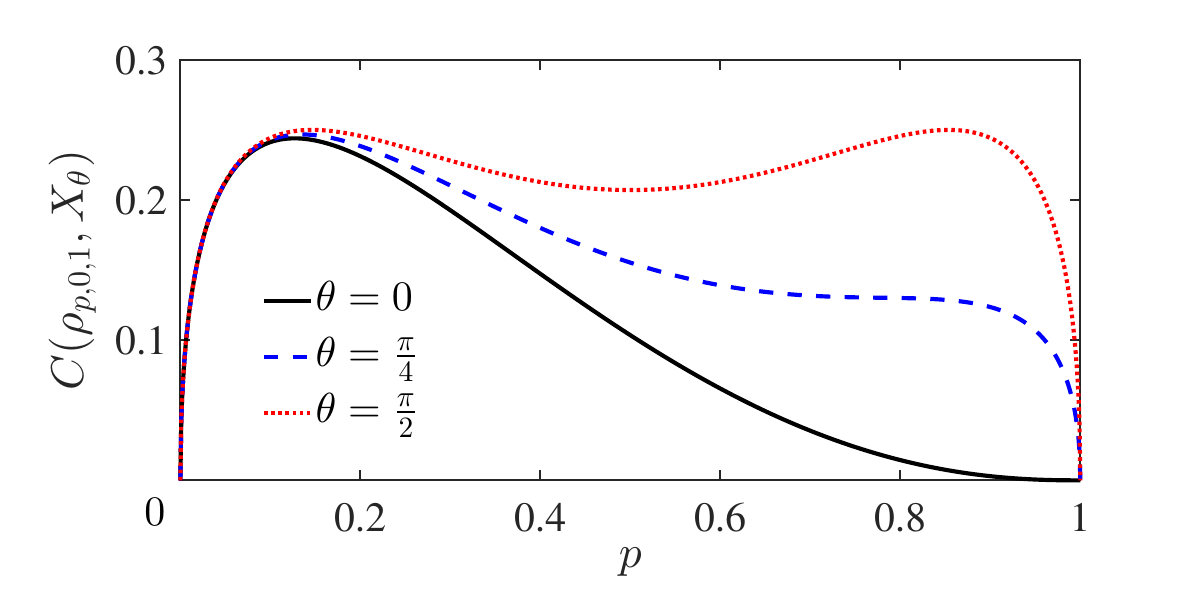}
}
 \caption{ Complexity of  state $\rho_{p,0,1}$  relative  to Hamiltonian $X_\theta$   versus  $p\in(0,1)$   when $\theta=0$ (black solid line), $\pi/4$  (blue dashed line)  and  $\pi/2$ (red dotted line). Concretely, $X_{0}=\sqrt{2}X$, $X_{\pi/4}=X+P$, $X_{\pi/2}=\sqrt{2}P$.}\label{fig7}
\end{figure}

For the mixed states $ \rho_{p,m,n}, \ m\neq n  $ defined by Eq. (\ref{rhopmn}), we have
\begin{align*}
&C(\rho_{p,0,n},X_\theta)=\Big(1-\frac{p^2\cos^2{\theta}}{2p^2-2p+1}\Big) f_p\delta_{n,1},\qquad \;  n>0,\\
&C(\rho_{p,1,n},X_\theta)=C(\rho_{p,n,1},X_\theta)=g_p/2,  \ \ \quad\quad \qquad n>1,\\
&C(\rho_{p,m,n},X_\theta)=0, \qquad \qquad \qquad\qquad \ \ \  n>1\& \ m>1,
\end{align*}
where $f_p$ and $g_p$  are defined by Eqs. (\ref{f}) and  (\ref{rhop})  respectively. The graph of $C(\rho_{p,0,1},X_\theta)$ versus the parameter $p$ is shown in Fig. \ref{fig7} when $\theta=0,\pi/4,\pi/2$. $C(\rho_{p,0,1},X_0)$ initially increases monotonically and then decreases in  $p$, reaching its maximum about $0.2440$ at $p\approx0.1269$.
$C(\rho_{p,0,1},X_{\pi/4})$ initially increases monotonically and then decreases in  $p$, reaching its maximum about $0.2468$ at $p\approx0.1351$.  When $\theta=\pi/2$, $C(\rho_{p,0,1},X_{\pi/2})=f_p$
 first increases monotonically in $[0,(2-\sqrt{2})/4)$, then decreases in $[(2-\sqrt{2})/4,1/2)$, then increases again in $[1/2,(2+\sqrt{2})/4)$, and finally decreases once more in $[(2+\sqrt{2})/4,1]$, and reaches its maximum $1/4$ at $p=(2\pm\sqrt{2})/4$. In particular,
\begin{align}\label{Cd}
C(\rho_{1/2,0,n},X_\theta)=\frac14(\sqrt{2}-1)(1+\sin^2{\theta})\delta_{n,1},
\end{align}
in sharp contrast to the result shown in Eq. (\ref{rho1/20n}).

%\begin{align*}
%&C(\rho_{p,0,n},X_\theta)\\&=\frac{(1-p)^2+p^2\sin^2{\theta}}{2p^2-2p+1}\sqrt{2p(1-p)}\Big(1-\sqrt{2p(1-p)}\Big)\delta_{n,1}.
%\end{align*}

%Untitled8-figure7

\vskip 0.2cm

3. {\sl Complexity relative to  $e^{-i\theta}a^2+e^{i\theta} a^{\dag2}$}

The Hamiltonian $\Lambda_{\theta}=e^{-i\theta}a^2+e^{i\theta} a^{\dag2},\ \theta\in[0,2\pi),$  is a quadratic operator.  The significance of $\Lambda_{\theta}$ lies in its ability to generate squeezing transformations, critical for nonclassical state preparation and quantum technologies. Here we illustrate the complexity quantifier  $C(\rho,\Lambda_\theta)$  relative to the Hamiltonian  $\Lambda_{\theta}$ by several  typical Bosonic states.

{\bf Proposition 6.}  

\noindent (1)  For any Fock-diagonal state $\rho_{\rm D},$ we have
\begin{align*}
C(\rho_{\rm D},\Lambda_\theta)=2 \sum_{n=0}^\infty \frac{ \lambda_{n}^{1/2}\lambda_{n+2}^{1/2}(\lambda_{n}^{1/2}-\lambda_{n+2}^{1/2})^2 }{\lambda_n+\lambda_{n+2}} (n+1)(n+2),
\end{align*}
which is independent of the parameter $\theta$.

\noindent (2)  It holds that 
\begin{align*}
C(D_z\rho_{\rm D} D_z^\dag,\Lambda_\theta)&=C(\rho_{\rm D},\Lambda_{\theta})+4|z|^2C(\rho_{\rm D},X_{\theta}),
\end{align*}
for any $\theta \in [0,2\pi).$

\noindent (3) It holds that 
%the amount of complexity for any squeezed  Fock-diagonal state $S_\zeta\rho_{\rm D} S_\zeta^\dag$  is
\begin{align*}
 C(S_\zeta\rho_{\rm D} S_\zeta^\dag,\Lambda_\theta)=|\beta^{'}_{\theta,\zeta}|^2 C(\rho_{\rm D},\Lambda_\theta),
\end{align*}
where
\begin{align}\label{betap}
\beta^{'}_{\theta,\zeta}=e^{i\theta}\cosh^2{|\zeta|}+e^{-i(\theta-2\arg\zeta)}\sinh^2|\zeta|.
 \end{align}

Next, we  consider some more specific examples.

For $\rho_{p,k}=p|0\rangle\langle0|+(1-p)|k\rangle\langle k|, \ k\geq1,$  we have
\begin{align*}
C(\rho_{p,k},\Lambda_\theta)=4\sqrt{p(1-p)}\big(1-2\sqrt{p(1-p)}\big)\delta_{k,2}=2g_p\delta_{k,2}.
\end{align*}
In particular, $C(\rho_{p,2},\Lambda_\theta)$  is twice the value of $C(\rho_p,X_\theta)$ as shown in Fig. \ref{fig4}.

For any thermal states $\tau_\lambda$ defined by Eq. (\ref{thermal}), we have
\begin{align*}
C(\tau_\lambda,\Lambda_\theta)=\frac{4\lambda}{1+\lambda^2},
\end{align*}
which monotonically increases in the  noise  parameter $\lambda$.

For the displaced thermal states $D_z \tau_\lambda D_z^\dag,$  we have
\begin{align*}
C(D_z \tau_\lambda D_z^\dag,\Lambda_\theta)=\frac{8\sqrt{\lambda}(1-\sqrt{\lambda})^2}{1-\lambda^2}|z|^2+\frac{4\lambda}{1+\lambda^2},
\end{align*}
which monotonically increases in  the displacement parameter $|z|$ for fixed $\lambda.$
%The graphs of $C(D_z \tau_\lambda D_z^\dag,\Lambda_\theta)$  versus $\lambda$  with fixed    $|z|=0.1,\,1,\,1.5,\,2$ are presented in Fig.  \ref{fig:11}.  This figure reveals that $C(D_z \tau_\lambda D_z^\dag,\Lambda_\theta)$  increases first and then decreases in  the noise parameter $\lambda$ for fixed  $|z|=0.1,\,1,\,1.5,\,2$.

For any squeezed thermal states $S_\zeta\tau_\lambda S_\zeta^\dag$, we have
\begin{align*}
 C(S_\zeta\tau_\lambda S_\zeta^\dag,\Lambda_\theta)= \frac{4\lambda }{1+\lambda^2}|\beta^{'}_{\theta,\zeta}|^2,
\end{align*}
where $\beta^{'}_{\theta,\zeta}$ is defined by Eq. (\ref{betap}).
%and $$|\beta^{'}_{\theta,\zeta}|^2=1+\sinh^2{(2|\zeta|)}\cos^2{(\theta-\arg\zeta)}.$$
The complexity $C(S_\zeta\tau_\lambda S_\zeta^\dag,\Lambda_\theta)$  is monotonically increasing in  the squeezing strength $|\zeta|$ for fixed $\lambda,\  \theta$,   and in the  parameter $\lambda$ for fixed $\zeta, \  \theta$. For fixed $\lambda$   and  $|\zeta|,$ $C(S_\zeta\tau_\lambda S_\zeta^\dag,\Lambda_\theta)$  is monotonically decreasing in  $\theta-\arg\zeta$ in $[k\pi,k\pi+\pi/2)$, and is monotonically increasing in $[k\pi+\pi/2,k\pi+\pi].$

\begin{figure}[htbp]
 \centering
{
 \includegraphics[width=0.48\textwidth]{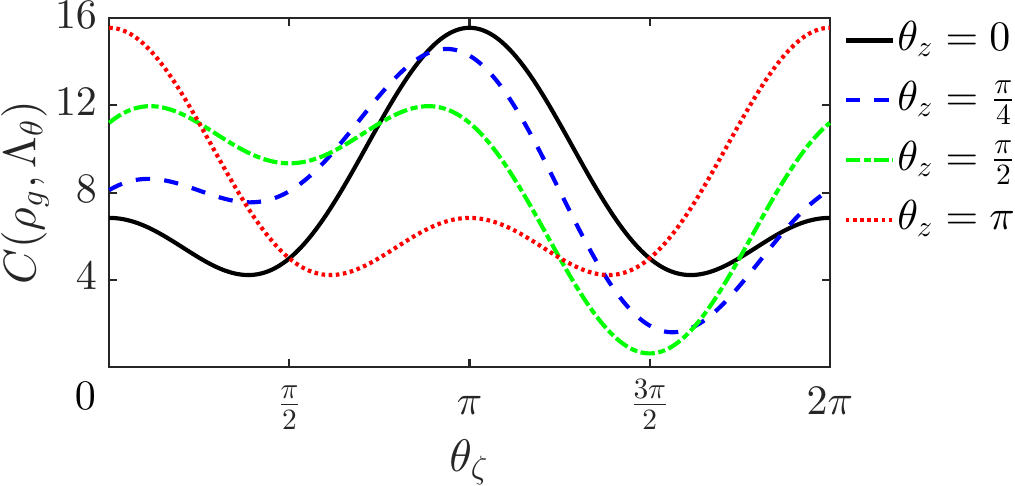}
}
 \caption{Complexity $C(\rho_{g},\Lambda_\theta)$ of Gaussian state $\rho_{g}$ relative  to  Hamiltonian $\Lambda_\theta$ versus   parameter   $\theta_\zeta=\theta-\arg{\zeta}$ for $\theta_z=0$ (black solid line), $\pi/4$ (blue dashed   line), $\pi/2$ (green dash-dotted line), and  $\pi$ (red dotted line)   with fixed  $|z|=1$,  $|\zeta|=1$ and  $\lambda=\lambda_0$. Here $\theta_z=\theta-2\arg z.$}\label{fig8}
\end{figure}

For the Gaussian states $\rho_g=D_z S_\zeta\tau_\lambda  S^{\dag}_\zeta D^\dag_z$, we have
\begin{align*}
C(\rho_g,\Lambda_\theta)=\frac{4\lambda}{1+\lambda^2}|\beta^{'}_{\theta,\zeta}|^2
+\frac{8\sqrt{\lambda}(1-\sqrt{\lambda})^2}{1-\lambda^2}|\beta_{\theta,z,\zeta}|^2,
\end{align*}
where $\beta^{'}_{\theta,\zeta}$ is defined by Eq. (\ref{betap}) and
\begin{align*}
&\beta_{\theta,z,\zeta}=z^*e^{i\theta}\cosh{|\zeta|}-ze^{-i\theta_\zeta}\sinh{|\zeta|},\\
&\theta_z=\theta-2\arg z,\\
&\theta_\zeta=\theta-\arg \zeta.
\end{align*}
For fixed  $\lambda$,  $\theta$  and $\zeta,$ $C(\rho_g,\Lambda_\theta)$  monotonically increases in  the displacement parameter $|z|$.
For fixed  $\lambda$,  $\theta$,    $z$ and $\arg\zeta,$ $C(\rho_g,\Lambda_\theta)$  monotonically increases in  the squeezing parameter $|\zeta|$ when $\cos{(\theta_z+\theta_\zeta)}\leq0$,
however monotonically decreases first and then increases  in  the squeezing parameter when $\cos(\theta_z+\theta_\zeta)>0.$ For fixed $\theta$, $z$  and $\zeta,$ $C(\rho_g,\Lambda_\theta)$  monotonically increases in  the noise parameter $\lambda$
when $\lambda\leq\lambda_0$ defined by Eq. (\ref{lambda0}). However, for $\lambda>\lambda_0$,  the monotonicity of $C(\rho_g,\Lambda_\theta)$  depends on the specific values of $\theta$,   $z$  and $\zeta$.
Moreover, as indicated in Fig.  \ref{fig8}, $C(\rho_g,\Lambda_\theta)$ does not have a consistent monotonicity in the parameter $\theta_\zeta$ with fixed  $|z|=|\zeta|=1$, $\lambda=\lambda_0$ and $\theta_z=0,\ \pi/4,\ \pi/2, \ \pi.$
%The graphs of $C(\rho_g,\Lambda_\theta)$  versus $\lambda$  with fixed  $|z|=2$, $(2\theta-2\arg z-\arg\zeta)=\pi/30$ and $\theta-\arg\zeta=\pi/4$ and $|\zeta|=0.01, \ 0.04,\  0.07, \ 0.1, \ 0.3, \ 0.5$ are shown in Fig.  \ref{fig8}. This figure demonstrates that  $C(\rho_g,\Lambda_\theta)$  initially  increases monotonically and then decreases in  $\lambda$ when the values of $\zeta$  and  $z$ are fixed at these specific numbers.   However, the maxmum point and its corresponding value vary depending on the parameters  $\theta$,    $z$ and  $\zeta$.
%untitled35-figure8

 For the mixed states $ \rho_{p,m,n}$ defined by Eq. (\ref{rhopmn}), when $m=0$ or $n=0,$ we have
\begin{align*}
C(\rho_{p,0,n},\Lambda_\theta)=C(\rho_{p,n,0},\Lambda_\theta)=\Big(2-\frac{2p^2\cos^2{\theta}}{2p^2-2p+1}\Big) f_p\delta_{n,2},
\end{align*}
and 
\begin{align*}
C(\rho_{p,n,m},\Lambda_\theta)=C(\rho_{p,m,n},\Lambda_\theta)=g_p\delta_{n,2},
\end{align*}
where   $f_p$  and $g_p$  are defined by  Eqs. (\ref{f}) and (\ref{rhop}),  respectively. In particular,
\begin{align}\label{C}
C(\rho_{1/2,0,n},\Lambda_\theta)=\frac14(\sqrt{2}-1)(3-\cos{2\theta})\delta_{n,2},
\end{align}
in sharp contrast to the result shown in Eqs. (\ref{rho1/20n}) and  (\ref{Cd}).
It is easy to observe the following relationship
\begin{align*}
C(\rho_{p,0,2},\Lambda_\theta)=2C(\rho_{p,0,1},X_\theta).
\end{align*}
The graph of $C(\rho_{p,0,1},X_\theta)$ is shown in Fig. \ref{fig7}.

\section{Comparison}
\begin{table*}[t]
    \centering
    \begin{tabular}{|c|c|c|}
     \hline
        Quantifier of complexity &Minimum achieved by&System  \\
        \hline
         $M_{\rm GFS}(\psi)$,  $M_{\rm MFS}(\psi)$   &Gaussian states& Pure, Continuous-variable  \\
         $C_{\rm T}(\rho)$  &(displaced) thermal states&Continuous-variable  \\
         $C(\rho,H)$ &pure and stable  states& Discrete or continuous-variable \\
          \hline
    \end{tabular}
    \caption{Comparison among three different quantifiers  of complexity}
    \label{tab:my_label}
\end{table*}
In this section,  we compare our quantifier of complexity  with those proposed by Manzano \cite{manzano2012statistical} and Tang {\it et al}. \cite{tang2025quantifying}.

Recall that the  quantifier of complexity  proposed by Manzano  is defined as the  product of Fisher information and  the  entropic power of the one-dimensional probability distribution for the measurement outcomes of an observable \cite{manzano2012statistical}
$$\tilde{X}_{t}=X\cos{t}-P\sin{t}=X_{-t}/\sqrt{2},\qquad t\in[0,\pi].$$  
Denoting the eigenvectors and eigenvalues of operator $\tilde{X}_{t}$ by $\{|x_{t}\rangle\}$ and $\{x_{t}\}$ respectively,  then the probability distribution of the measurement of  $\tilde{X}_{t}$  in the  pure state $|\psi\rangle$ is  
$q_t (x_{t})=|\langle x_t|\psi\rangle|^2,$ and the statistical complexity proposed by Manzano reads \cite{manzano2012statistical}
\begin{align*}
M_{\rm FS}(\psi|t)=I (q_t \big)J\big(q_{t}),
\end{align*}
where $I\big(q_{t})$ is the Fisher information of $q_t$ with respect to the location, i.e., of $p_\theta (x)=q_t(x-\theta)$ as defined by Eq. (\ref{D1}), and
\begin{equation*}
J(q_{t})=\frac{1}{2\pi e}e^{2H (q_{t})}
\end{equation*}
is the entropic power of Shannon
entropy $H (q_{t})=-\int^{\infty}_{-\infty}q_t (x_{t})\ln{q_t (x_{t})}{\rm d}x_{t}.$  $M_{\rm FS}(\psi|t)$ depends on the basis of the measurement, in other words, its dependence on the parameter $t$ is non-trivial. Manzano obtained two quantifiers independent of the measurement basis by averaging and minimizing  over  the parameter $t,$
 \begin{align*}
&M_{\rm GFS}(\psi)=\frac{1}{\pi}\int^{\pi}_{0}M_{\rm FS}(\psi|t){\rm d}t,\\
&M_{\rm MFS}(\psi)=\min_{t\in[0,\pi]}M_{\rm FS}(\psi|t),
\end{align*}
which are  called global Fisher-Shannon quantifier and minimum Fisher-Shannon quantifier of complexity, respectively. Both quantifiers are limited to pure states and reach their minimum values for Gaussian states.

The  quantifier  of complexity proposed by Tang  {\it et al}  is the product of Fisher information and Wehrl entropy power based on  the Husimi quasiprobability  distribution $Q(\alpha|\rho)=\langle\alpha|\rho|\alpha\rangle$ of any single-mode bosonic  state $\rho$ \cite{tang2025quantifying}, that is 
\begin{align*}
C_{\rm T}(\rho)=e^{S_{\rm W}(\rho)-1}I_{\rm F}(\rho),
\end{align*}
where 
\begin{align*}
&I_{\rm F}(\rho)=\frac{1}{4}\int_{\mathbb{C}}\frac{||\nabla Q(\alpha|\rho)||^2}{Q(\alpha|\rho)}\frac{{\rm d}^2\alpha}{\pi},\\
&S_{\rm W}(\rho)=-\int_{\mathbb{C}}Q(\alpha|\rho)\ln{Q(\alpha|\rho)}\frac{{\rm d}^2\alpha}{\pi},
\end{align*}
which are the  (classical) Fisher information of the Husimi function  and  the Wehrl entropy,  respectively. Here ${\rm d}^2\alpha ={\rm d}x{\rm d}y$ (with $\alpha =x+iy$) is the Lebesgue measure on $\mathbb{C}$.  The gradient operator $\nabla $ is   taken  with respect to the real and imaginary parts of the
complex number $\alpha$.
The Wehrl entropy $S_{\rm W}(\rho)$  represents the spread of $\rho$, while the Fisher
information $I_{\rm F}(\rho)$  represents the localization of $\rho$ in  phase space. The complexity quantifier $C_{\rm T}(\rho)$ possesses some favorable properties, such as invariance under displacements and phase-space rotations, as well as under uniform phase-space scaling.  The quantifier $C_{\rm T}(\rho)$  reaches its minimum value only for the (displaced)  thermal  states.  
%The Husimi distribution is a phase-space distribution and can be regarded as a two-dimensional probability distribution \cite{Husimi}.

The quantifier of  complexity  we proposed is defined as the difference between  two important versions of quantum Fisher information: the quantum Fisher
information defined via the symmetric logarithmic derivatives and the Wigner-Yanase skew information, and incorporates  the Hamiltonian to reflect the dynamic variation of quantum states. Unlike the quantifiers proposed by Manzano and Tang {\sl et al}, our quantifier  is not limited to continuous-variable quantum systems.
It inherits several favorable properties of quantum Fisher information, such as additivity in the sense of  Eq.  (\ref{additive}). 

Another important difference among the three complexities lies in the states with the lowest complexity. The quantifiers $M_{\rm GFS}(\psi)$ and   $M_{\rm MFS}(\psi)$  of Manzano      are based on the one-dimensional probability distribution obtained through measurement.  In the one-dimensional
variable case, Gaussian states have normal distributions, which saturate the isoperimetric inequality.  Consequently, Gaussian states minimize Manzano's complexity. The quantifier $C_{\rm T}(\rho)$  is  based on a two-dimensional probability distribution (i.e., Husimi  quasiprobability distribution \cite{Husimi}). In the two-dimensional variable case, the (displaced) thermal states (without squeezing) follow normal distributions which also saturate the relevant isoperimetric inequality, leading them to minimize $C_{\rm T}(\rho)$ \cite{tang2025quantifying}.  Our quantifier is based on the fact that quantum dynamics  vanishes when the  states are pure states or  stable states, and  can distinguish different Gaussian states.  
%but the other two quantifiers cannot.

We  summarize the comparisons of the three quantifiers of complexity in Table \ref{tab:my_label}.

\section{Summary}
In this work,  we  have proposed a quantifier of  complexity for  quantum states based on  Fisher discord.  We have revealed some  properties of this complexity quantifier.
To illustrate the complexity quantifier, we have evaluated it in both discrete and continuous-variable quantum systems for some paradigmatic states.
% This Hamiltonian-dependent approach is a significant innovation, reflecting the dynamic behavior of quantum states  under Hamiltonian-driven evolution and providing deep insights into quantum dynamics.   %We expect to discover its potential applications in quantum metrology and quantum dynamics.

We have compared our complexity quantifier with the Fisher-Shannon quantifier and the Fisher-Wehrl quantifier in the literature. Our findings highlight notable distinctions between these approaches. These quantifiers  assess complexity from  distinct perspectives, offering distinct strengths in various contexts. Further comprehensive studies comparing other complexity quantifiers would be highly beneficial.

Gibilisco {\it et al}. proved that the metric-adjusted skew information is dominated by the quantum  Fisher information based on the symmetric logarithmic derivative \cite{gibilisco2009inequalities}.  It is also interesting to study the difference between these two (generalized) quantum Fisher information, which  is a natural generalization of Fisher discord.

Both the quantum Fisher information based on the symmetric logarithmic derivative and the skew information  does not increase under the partial traces, that is,
\begin{align*}
   I_{\rm F}(\rho_{12},H\otimes{\bf 1})&\geq I_{\rm F}(\rho_{1},H),\\
   I_{\rm W}(\rho_{12},H\otimes{\bf 1})&\geq I_{\rm W}(\rho_{1},H),
\end{align*}
where $\rho_{12}$ is a bipartite state of the composite system, $\rho_{1}={\rm tr}_2\rho_{12}$ is the reduced state of system 1 and $H$ is an observable on system 1. It is natural to consider such an open question: Whether quantum Fisher discord $$C(\rho_{12},H\otimes{\bf 1})=I_{\rm F}(\rho_{12},H\otimes{\bf 1})-I_{\rm W}(\rho_{12},H\otimes{\bf 1})$$ for  bipartite states does not increase under the partial trace. If the answer is affirmative, it is worth investigating  correlations of  the bipartite states  via the difference as
$C(\rho_{12},H\otimes{\bf 1})-C(\rho_{1},H).$

Given the significance of quantum Fisher information in metrology and information theory, there is compelling motivation to investigate the theoretical implications and practical applications of Fisher discord further. Applying these quantifiers to study  correlations, interference, quantum metrology, and coherence holds significant promise, warranting deeper exploration. 

\vskip 0.4cm
\noindent {\bf Acknowledgements}.
 This work was supported by the National Key R\&D Program of China, Grant No. 2020YFA0712700, the National Natural Science  Foundation of China, Grant Nos. 12401609, 12426671 and 12341103, the Youth Innovation Promotion Association of CAS, Grant No. 2023004, and the Shuimu  Tsinghua
Scholar Program of Tsinghua University and  the Postdoctoral Science Foundation of China, Grant No.  GZB20250711.

%\bibliography{reference}
%\bibliographystyle{aapmrev4-2}
%\bibliographystyle{apsrev4-2}
\end{document}